\documentclass[aps,prl,groupedaddress,fleqn,twocolumn]{revtex4}
\pdfoutput=1
\usepackage{graphicx}
\usepackage{amsmath}
\usepackage{gensymb}

\begin{document}

\title{Slow stretched-exponential and fast compressed-exponential relaxation from local event dynamics}
\author{K. Trachenko$^{1}$}
\author{A. Zaccone$^{2}$$^{,}$$^{3}$}
\address{$^1$School of Physics and Astronomy, Queen Mary University of London, Mile End Road, London, E1 4NS, UK}
\address{$^2$Department of Physics ``A. Pontremoli'', University of Milan}
\address{$^3$Cavendish Laboratory, University of Cambridge, CB3 0HE, Cambridge, U.K.}

\begin{abstract}
We propose an atomistic model for correlated particle dynamics in liquids and glasses predicting both slow stretched-exponential relaxation (SER) and fast compressed-exponential relaxation (CER). The model is based on the key concept of elastically interacting local relaxation events. SER is related to slowing down of dynamics of local relaxation events as a result of this interaction, whereas CER is related to the avalanche-like dynamics in the low-temperature glass state. The model predicts temperature dependence of SER and CER seen experimentally and recovers the simple, Debye, exponential decay at high temperature. Finally, we reproduce SER to CER crossover across the glass transition recently observed in metallic glasses. %The proposed picture unifies different atomistic dynamics and relates it to the underlying energy landscape morphology and experimental phenomena.
\end{abstract}

%\pacs{61.43.Fs, 64.70.Pf, 61.20.Lc}

\maketitle

\section{Introduction}

As a matter of general principle in physics, correlations tend to decay in time and space. Persisting currents are an exception and operate in special cases such as superconductivity or superfluidity. The most common time decay function is the exponential decay:
\begin{equation}
\phi=\phi_0e^{-\frac{t}{\tau}}
\end{equation}

\noindent where $\tau$ is relaxation time which quantifies the time of return to equilibrium.
Eq. (1) is the hallmark of linear dynamics in equilibrium systems.

In disordered systems such as glasses, low-temperature viscous liquids and biological systems, time decay is notably different and is commonly described by stretched-exponential relation (see, e.g., Refs. \cite{dyre,fuchs,voigtmann,ngai,wolynes,schweizer,phillips1,phillips2,brujic,Zaccone_review}):
\begin{equation}
\phi=\phi_0e^{-\left(\frac{t}{\tau}\right)^\beta}
\end{equation}

\noindent where $0<\beta<1$.

Stretched-exponential relaxation (SER) describes slow dynamics: in an intermediate data range, the decay of $\phi$ can be approximated as a logarithm of time as seen experimentally \cite{bra}.

Since SER was introduced by Kohlrausch in 1854 \cite{kohl}, understanding it has remained one of the oldest problems in physics \cite{phillips1,phillips2}. Several mechanisms of SER have been proposed. One class of models involves axiomatic theories of SER involving sinks and traps \cite{phillips1,phillips2}. Another popular class assumes that a disordered system has a distribution of local relaxation times or local activation barriers, and aims to derive SER as an integral over the distribution (see, e.g., Ref. \cite{palmer}). Depending on model details, this approach can reproduce the asymptotic behavior of SER at long waiting times but not in the range where SER varies most \cite{zwanzig}. A different approach \cite{prb} puts an emphasis on non-ergodic irreversible nature of relaxation where activation barriers evolve during the relaxation process. In this process, an activation barrier of a local relaxation event (LRE) is set by previous LREs through a feed-forward elastic interaction mechanism.

More recently, a surprising result emerged \cite{ruta} showing that $\beta$ can be larger than 1 in supercooled liquids below the glass transition, prompting the term ``compressed exponential relaxation'' (CER). This is an intriguing result showing that relaxation can be faster than exponential, Debye, relaxation. Furthermore, a crossover from SER to CER has been experimentally observed across the glass transition~\cite{ruta,kob}, while CER has also been observed in soft glasses~\cite{Egelhaaf}. There is currently no theoretical explanation of these effects in general terms. A theoretical model for colloidal gels gives $\beta=3/2$ \cite{Bouchaud}. Experimentally, $\beta$ varies with temperature, and a continuous spectrum of $\beta>1$ values in CER have been measured \cite{ruta,betacer}. Hence of particular importance is understanding (a) the microscopic mechanism of CER on the basis of a generic mechanism and its temperature dependence and (b) the mechanism of the crossover from SER to CER.

In this paper, we show that CER follows from a kinetic equation governing the dynamics of elastically interacting LREs. This involves two main steps. First, we discuss the key physical process related to the cooperative, non-independent, nature of relaxation of local events. In equilibrium viscous liquids, the cooperativity arises from the re-distribution of local stresses and resulting elastic interaction between LREs. This slows down the LRE dynamics and gives SER as a solution of the rate equation for LREs in the supercooled liquid. Second, we observe that relaxation in highly non-equilibrium systems below the glass transition temperature $T_{g}$ is related to the avalanche-like dynamics \cite{Evenson,egami}. Assuming that the relaxation of later LREs is promoted in this process, we write the corresponding rate equation and show that it gives CER. The two processes can be combined into a unifying description that provides the experimentally observed crossover from SER above $T_g$ to CER below $T_g$.

\section{Elastic interaction between local relaxation events}

\subsection{Concordant relaxation and stress redistribution}

In this and next section, we discuss the mechanism from which either SER or CER follows depending on the nature of interaction between LREs.

Let us consider a low-temperature viscous liquid or glass under fixed external perturbation such as constant shear or compressive stress. In the case of compressive stress, the response of a relaxing system is the sum of the viscous and elastic components \cite{frenkel}. The viscous component discussed here decays to zero after the relaxation is complete, and the elastic component remains. Notably, LREs are not independent but elastically interact as discussed below.

Earlier work considered local relaxation events in glasses and related elastic effects \cite{argon1,argon2,orowan,deng,bulatov}. Under sufficiently high stress, each LRE in glasses can involve an atom leaving its surrounding cage with associated bond-breaking and bond-forming atoms in the nearby local environment \cite{rebond,rebond1}. Orowan introduced a ``concordant'' LRE accompanied by a strain agreeing in direction with the stress field \cite{orowan} and reducing the energy and local stress. This has led to the result that stress relaxation by earlier concordant events leads to the increase of stress on later relaxing regions in a system. Goldstein applied the same argument to a viscous liquid \cite{gold}: consider a system under external stress. Initially, the external stress is counter-balanced by a network of stress-supporting local regions. When a local concordant rearrangement to a potential minimum, biased by the external stress, occurs, this local region supports less stress after the event than before; therefore, other local regions in the vicinity should support more stress after that event than before (in this process, nearby regions are affected more by stress-redistribution due to stress decay) \cite{gold}. Goldstein proposed that ``the least any model of the flow process must acknowledge is that the extra stress must be supported elsewhere''.

Each LRE carries a microscopic change of a macroscopic quantity (e.g., microscopic stress). Consequently, the number of events governs the dynamics of a relaxing macroscopic observable. Let us consider the current number of LREs $n(t)$, induced in a system by an external perturbation such as fixed shear or compressive stress or by a long-range internal stress field. This number $n(t)$ comes in addition to thermally-induced LREs. When $n(t)$ tends to its limiting value $n_r$ at long times, the perturbation is relaxed to zero.

As discussed by Orowan and Goldstein, because the perturbation field introduces bias towards concordant relaxation events supporting less stress after relaxation, later LREs should support more stress in order to counterbalance. Therefore, the increase of stress on a local region, $\Delta p$, increases with $n$. This affects the activation barrier for a LRE, $V$, which is governed by the elastic energy \cite{dyre,frenkel}. This picture is consistent with glassy relaxation in supercooled liquids, where the system trickles down towards lower states in the energy landscape, characterized by higher barriers~\cite{Ramos}.

\subsection{Feed-forward interaction mechanism}

At a fixed cage volume, the energy needed for the central atom to leave the cage is very high due to strong interatomic repulsion, resulting in very long waiting times. On the other hand, a temporary increase of the cage volume (e.g. due to thermal fluctuations) promotes the LREs \cite{frenkel}. However, this increase is opposed by elasticity of the surrounding system. Therefore, Frenkel calculates $V$ as the work done to expand the cage in order for a LRE to take place \cite{frenkel}. This work changes as a result of the increase of stress on a local region due to stress redistribution as discussed in the previous section. This, in turn, changes $V$ as discussed below.

Loading additional stress $\Delta p$ on the current local region increases $V$. In the case of external shear stress, this is seen by noting that $V$ is set by the stored elastic shear energy in the surrounding of the local region as discussed in the shoving model \cite{dyre}. Since the additional shear stress $\Delta p$ increases the elastic shear energy, $V$ increases. The same applies to the case of external compressive stress: loading additional compressive stress $\Delta p$ increases the work needed to expand the cage and, therefore, $V$. In either case, the increase of $V$ is given by the increase of work needed to overcome the additional barrier created by the elastic force due to the additional stress $\Delta p$ as $\Delta V=\int \Delta p {\rm d}q$, where $q$ is the cage volume. If $q_0$ is a characteristic cage volume, $\Delta V=\Delta p q_0$, giving

\begin{equation}
V(n)=V_0+q_0\Delta p
\label{vn0}
\end{equation}

\noindent where $V_0$ is the initial value of the barrier.

Eq. \eqref{vn0}, together with $\Delta p$ increasing with $n$ as discussed above, imply that $V(n)$ increases with $n$. This describes the {\it feed-forward interaction mechanism} between LREs, in that activation barriers increase for later events. We note that this mechanism is related to the ability of liquids to support stress, a property ascertained in theory and experiments \cite{ropp,noirez1,noirez2,PNAS2020}. The mechanism is similar in spirit to dynamic facilitation in kinetically constrained models of liquid-glass transition where relaxation of one event is affected by others \cite{chandler,garrahan,harrowell}.

We now introduce a dynamical variable $n(t)$, the current number of LREs in a sphere of radius $d_{el}$, where $d_{el}$ is the elasticity length $d_{el}=c\tau$, the length over which stress propagates in the liquid \cite{del,ropp} (here $c$ is the speed of sound and $\tau$ is relaxation time). $n(t)$ starts from zero and increases to its final value $n_r$: $n(t)\rightarrow n_r$ at long times. Let us consider the current LRE, about to relax, to be in the centre of the sphere. All previous concordant LREs that are located within distance $d_{el}$ from the centre, participate in the feed-forward interaction, increasing stress on the central region and hence the activation barrier for the central LRE.

Let $\Delta p_i(0)$ be the reduction of local stress due to a remote concordant LRE $i$. Then, $\Delta p_i$ decays with distance, hence we denote $\Delta p_i(r)$ as its value at distance $r$ from the centre. We assume, for simplicity, that $\Delta p_i(0)$ are constant, $\Delta p_i(0)=\Delta p_0$. The increase of stress on the central rearranging region, $\Delta p$, can be calculated by integrating $\Delta p_i(r)$:

\begin{equation}
\Delta p=\rho\int\limits_{d_0/2}^{d_{el}} 4\pi r^2 \Delta p_i (r) {\rm d}r
\label{int}
\end{equation}
where $d_0$ is on the order of the size of a local region and $\rho$ is the density of LREs, $\rho=6n/(\pi d_0^3)$.

In elasticity, stresses decay as $\Delta p(r)\propto \frac{1}{r^3}$ \cite{elast}. Since $\Delta p(r)=\Delta p_0$ at $r=d_0/2$, $\Delta p(r)=\Delta p_0(d_0/2r)^3$. Integrating Eq. (\ref{int}) and combining it with Eq. (\ref{vn0}) gives

\begin{eqnarray}
\begin{split}
& V=V_0+V_1\frac{n}{n_r}\\
& V_1=\frac{\pi}{2}\rho_r q_0\Delta p_0 d_0^3\ln\left(\frac{2d_{el}}{d_0}\right)
\end{split}
\label{vn}
\end{eqnarray}
\noindent where $\rho_r=6n_r/(\pi d_0^3)$ is the density of the final number of events in the sphere. We assume small external perturbations (small external stress and, consequently, small $\Delta p_0$), resulting in $V_1<V_0$.

\section{Exponential, stretched and compressed relaxation from the rate equation for LREs}

We are now set to derive both SER and CER as well as simple exponential relaxation on the basis of LRE dynamics. The rate of LREs, ${\rm d}n/{\rm d}t$, is proportional to the number of unrelaxed events, $(n_{\rm r}-n)$, and the event probability, $\rho=\exp\left(-\frac{V}{T}\right)$ ($k_{\rm B}=1$). Introducing $q=n/n_{\rm r}$, we write:

\begin{equation}
\frac{{\rm d}q}{{\rm d}t}=\exp\left(-\frac{V_{0}}{T}\right)(1-q) \exp(-\alpha q)
\label{rate}
\end{equation}
\noindent where $\alpha=\frac{V_1}{T}$, $t$ is dimensionless time $t=\frac{t}{\tau_0}$ and $\tau_0$ is the characteristic time.

When $d_{el}$ is small at high temperature, and equal to half of the cage size, $d_0/2$, $V_1=0$ in \eqref{vn}. This implies no elastic interactions between LREs because the this interaction does not propagate beyond nearest neighbours. Then, $\alpha=0$ in \eqref{rate}, and the rate equation is

\begin{equation}
\frac{{\rm d}q}{{\rm d}t}=\exp\left(-\frac{V_0}{T}\right)(1-q)
\label{exp}
\end{equation}

Equation \eqref{exp} results in the exponential relaxation $q\equiv n/n_{\rm r}=1-\exp(-{t/\tau})$, where $\tau=\tau_0\exp(V_0/T)$. This is consistent with SER becoming exponential at high temperature \cite{dyre,ngai}. At low temperature, \eqref{rate} gives SER as discussed below.

Therefore, our theory predicts a physically transparent {\it crossover} from SER at low temperature to simple exponential at high as is seen experimentally. The crossover corresponds to $d_{el}=c\tau=d_0$, giving $\tau$ at the crossover as $\tau=\frac{d_0}{c}$. Taking $d_0\approx 10$~\AA\ and $c\approx$1000 m/s gives $\tau$ at the crossover of about 1 ps, in agreement with several experiments \cite{del1,del2,del3,del}.

At low temperature, $d_{el}>d_0$, $V_1\ne 0$ in \eqref{vn} and $\alpha\ne 0$ in \eqref{rate}. Consequently, the relaxation governed by Eq. \eqref{rate} becomes non-exponential. We note that $d_{el}$ reaches system size $L$ at the dynamical crossover corresponding to liquid relaxation time $\tau=\frac{L}{c}$ of about 10$^{-6}$ s for $L$ typically $L\approx 1$ mm \cite{del}. This takes place well above the glass transition temperature. Hence, $V_1$ in \eqref{vn} becomes temperature-independent at low temperature when $d_{el}=L$ : $V_1=\frac{\pi}{2}\rho_r q_0\Delta p_0 d_0^3\ln\left(\frac{2L}{d_0}\right)$, and we will assume below that $V_1$ is constant. We note that this picture predicts (a) a crossover of $\beta$ at $d_{el}=L$ at low temperature where $\tau$ increases with $L$ as a power law due to Eq. \eqref{vn} \cite{del}, in agreement with experimental data \cite{size} (a caveat here is that whereas $d_{el}=c\tau$ describes the dominant mechanism of wave dissipation in liquids, other mechanisms are present as well, including the usual interatomic potential anharmonicity present in solids, the disordered structure and so on \cite{diss} and, consequently, the experimental size effect tends to be weaker as compared to theory \cite{size}), (b) less sensitive temperature dependence of $\beta$ after the crossover at low temperature because $V_1$ become constant, consistent with experimental results \cite{betac1,betac2,betac3} and (c) the decrease of $\beta$ with $L$ at low temperature after the crossover because $\tau$ and $\beta$ are anti-correlated \cite{betac1}.

We integrate Eq. \eqref{rate} numerically and, using the least-squares method, fit the result to

\begin{equation}
q=1-e^{-\left(\frac{t}{\tau}\right)^\beta}
\label{qfit}
\end{equation}

\noindent reflecting the fact that all LREs have taken place at long times and the fraction of LREs $q=\frac{n}{n_r}\rightarrow 1$ for $t\gg\tau$.

We plot $q$ at two different temperatures in Fig. 1a and observe a good agreement with SER and $\beta<1$. Lowering the temperature decreases $\beta$, consistent with experimental results \cite{dyre,ngai}.

\begin{figure}
{\scalebox{0.36}{\includegraphics{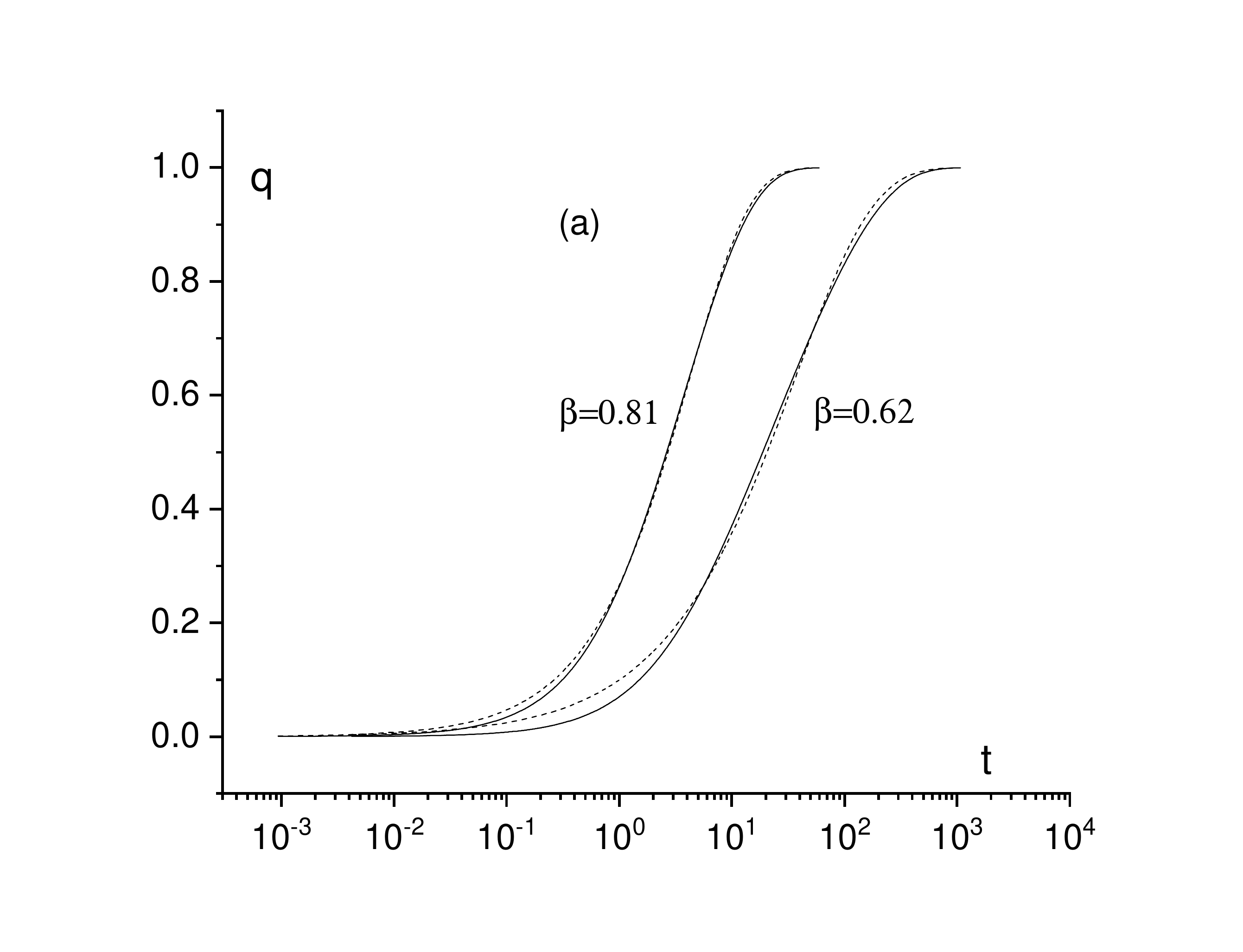}}}
{\scalebox{0.36}{\includegraphics{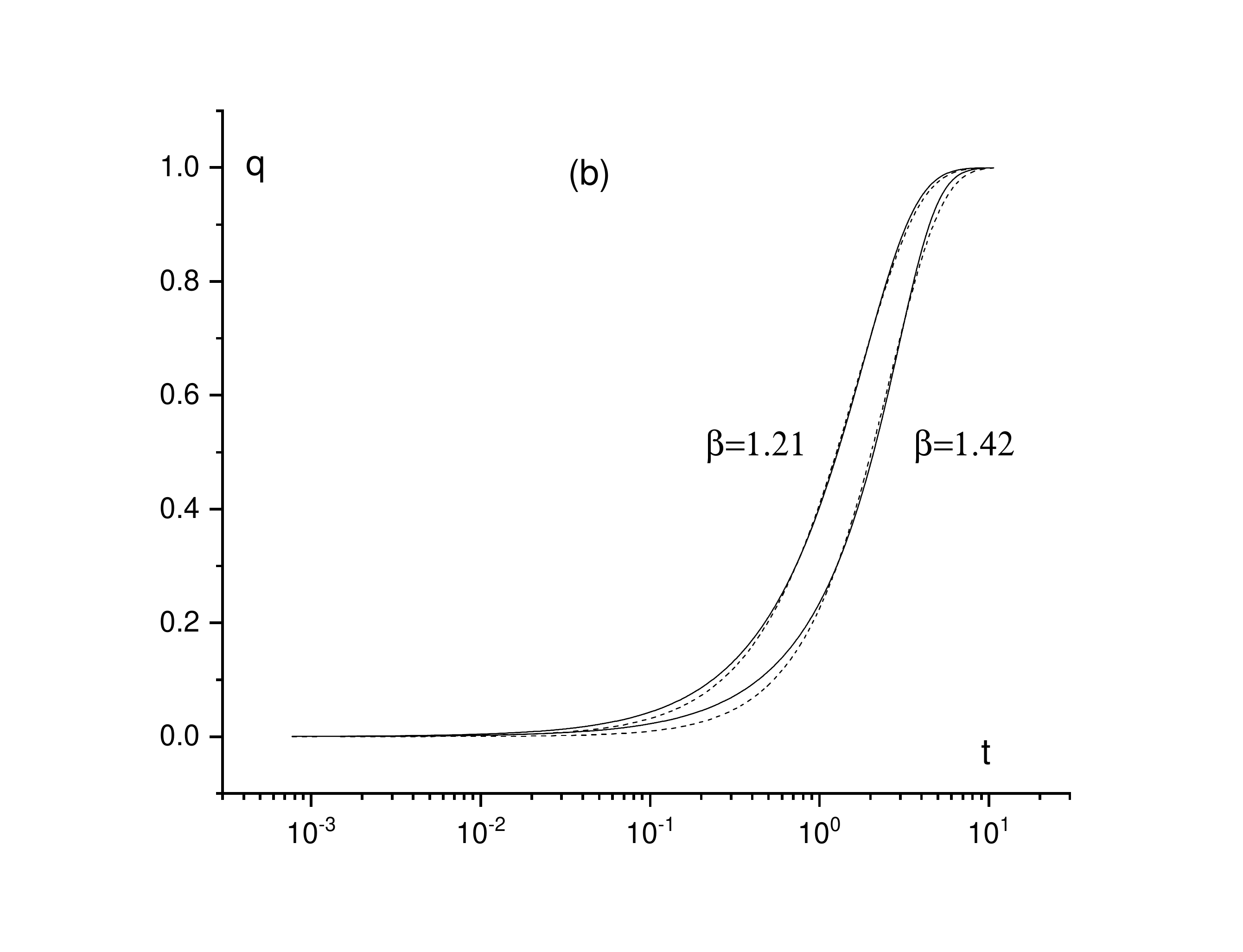}}}
\caption{(a) Solid lines show the solution of the SER equation \eqref{rate} (solid line) at two different temperatures corresponding to $\frac{V_0}{kT}=\frac{V_1}{kT}=\alpha\approx 1$ and $2.5$ ($V_0=V_1$ is set for simplicity and illustration purposes). Dashed lines show the fit of the solutions to Eq. \eqref{qfit} with the fitted values ($\beta=0.81$, $\tau=4.04$) and ($\beta=0.62$, $\tau=33.9$). (b) Solid lines show the solution of the CER equation \eqref{rate1} at two different temperatures corresponding to $\frac{V_0}{kT}=\frac{V_1}{kT}=\alpha\approx 0.8$ and $1.5$. Dashed lines show the fit of the solution to \eqref{qfit} with ($\beta=1.21$, $\tau=1.67)$ and ($\beta=1.42$, $\tau=2.62$).}
\label{fig1}
\end{figure}

We now discuss the mechanism of CER. The Orowan-Goldstein stress redistribution mechanism assumes that elastic quasi-equilibrium is established after each LRE. On the other hand, in systems such as metallic glasses where CER is seen, relaxation below the glass transition temperature is suggested to proceed via avalanches of local events \cite{Evenson,egami}. In this process, LREs are considered to take place in highly non-equilibrium stressed states where one LRE weakens its surrounding and promotes other LREs. This implies the reduction of activation barriers for later events, the opposite of the Orowan-Goldstein mechanism, and can therefore be represented as $V=V_0-V_1^\prime\frac{n}{n_r}$
($V_1^\prime<V_0$ as before), and with the sign in front of the time-dependent term $\frac{n}{n_r}$ opposite to that in \eqref{vn}. Then, the rate equation becomes

\begin{equation}
\frac{{\rm d}q}{{\rm d}t}=\exp\left(-\frac{V_{0}}{T}\right)(1-q)\exp(\alpha^\prime q)
\label{rate1}
\end{equation}
\noindent where $\alpha^\prime=\frac{V_1^\prime}{T}$ and, differently to \eqref{rate}, the sign in the exponent $\exp(\alpha^\prime q)$ is positive.

We integrate \eqref{rate1} numerically for different temperatures, assuming $V_1=V_1^\prime$ and $\alpha=\alpha^\prime$ for simplicity and for the purpose of comparing SER and CER. The result of integration is shown in Fig. 1b. We fit $q$ to Eq. \eqref{qfit} and observe a good agreement with CER with $\beta>1$. This is seen in Fig. 1b where we plot $q$ at different temperatures. Lower temperature increases $\beta$, consistent with experimental results \cite{betacer}.

In Fig. 2 we plot $\beta$ for both SER and CER in a wider range of temperature and observe that $\beta$ decreases for SER and increases for CER away from 1. The decrease of $\beta$ in the SER regime at low temperature is consistent with experiments \cite{dyre,ngai} as mentioned earlier. In contrast, we observe that $\beta$ increases in the CER regime at low temperature. This is also consistent with the experimental data \cite{betacer}.

\begin{figure}
{\scalebox{0.36}{\includegraphics{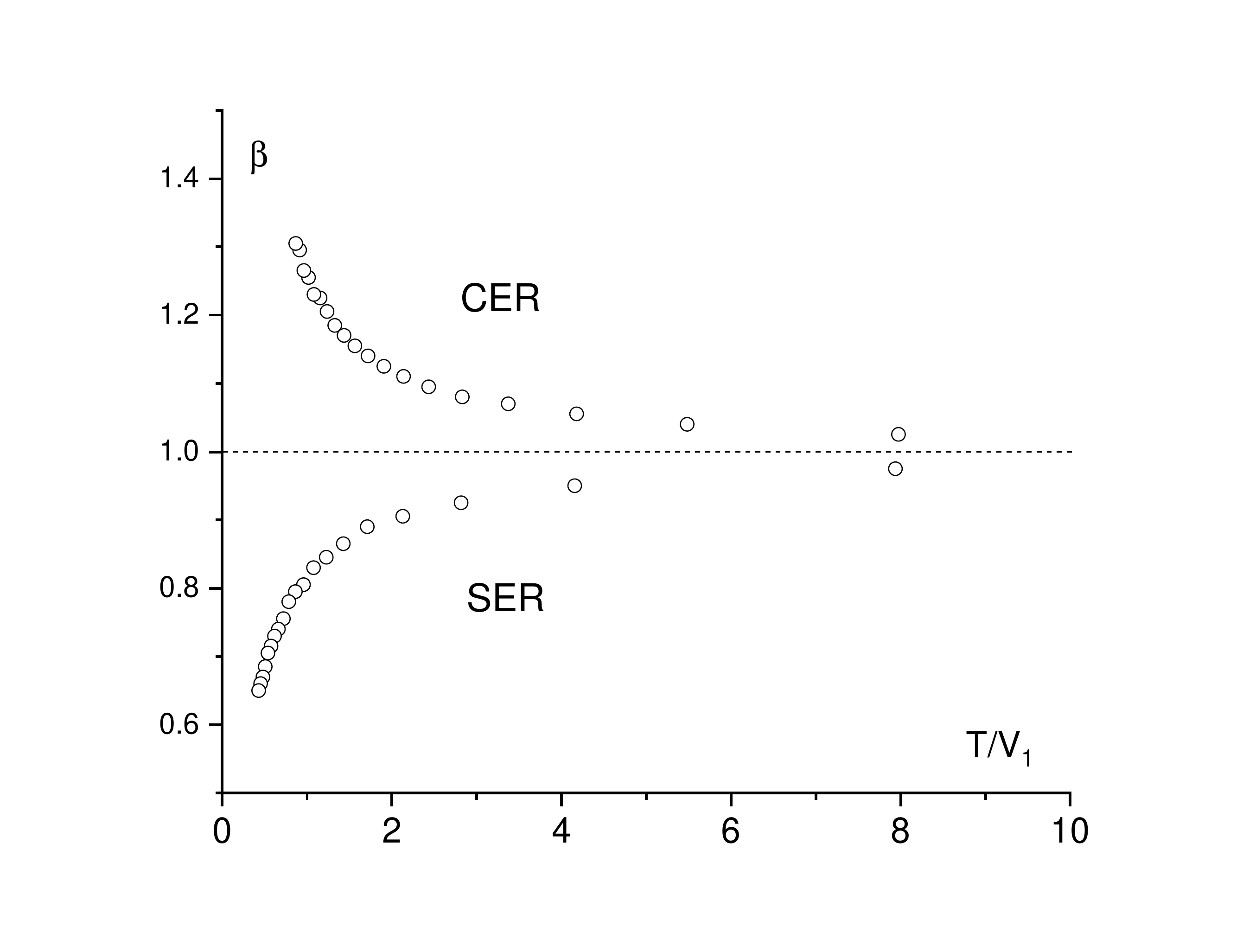}}}
\caption{$\beta$ for stretched-exponential and compressed-exponential relaxation as a function of reduced temperature $\frac{T}{V_1}$, obtained from solving Eq. \eqref{rate} for SER and Eq. \eqref{rate1} for CER and fitting the results to Eq. \eqref{rate}.}
\label{fig2}
\end{figure}

\section{SER-CER crossover}

We now demonstrate that our model can reproduce the crossover from SER to CER. As discussed above, the feed-forward ``slow-down'' interaction between LREs operates in the liquid above the glass transition temperature $T_g$, whereas the avalanche-like LREs take place at and below $T_g$. We can therefore assume that around $T_g$, both processes operate, but the intensity of the first and second process decreases and increases, respectively, as $T_g$ is approached from above. The activation energy barriers in the two processes depend on the relative number of LREs as $V_{SER}=V_0+V_1q$ and $V_{CER}=V_0-V_1q$ as discussed above. Therefore, we write the rate equation as
\begin{eqnarray}
\begin{split}
& \frac{{\rm d}q}{{\rm d}t}=(1-q)\exp\left(-\frac{V}{T}\right),\\
& V=(1-f)V_{SER}+fV_{CER},\\
& V_{SER}=V_0+V_1q; ~~~ V_{CER}=V_0-V_1q,
\end{split}
\label{inter}
\end{eqnarray}

\noindent where function $(1-f)$ defines the fraction of SER processes related to the Orowan-Goldstein mechanism and $f$ defines the fraction of CER processes related to avalanches in the non-equilibrium glass structure.

$V$ in \eqref{inter} reflects the change in the system's energy landscape, from SER- to CER-dominated. The purpose of the function $f$ is to vary between 0 at high $T$ and 1 at low $T$ to result in the SER-CER crossover. A convenient choice is to use the analogy with thermal two-level systems and represent $f$ as $f=\frac{1}{e^{-A\left(\frac{1}{T}-\frac{1}{T_0}\right)}+1}$, where $A$ and $T_0$ are parameters. $f$ is shown in Fig. 3 as a dashed line.

\begin{figure}
{\scalebox{0.36}{\includegraphics{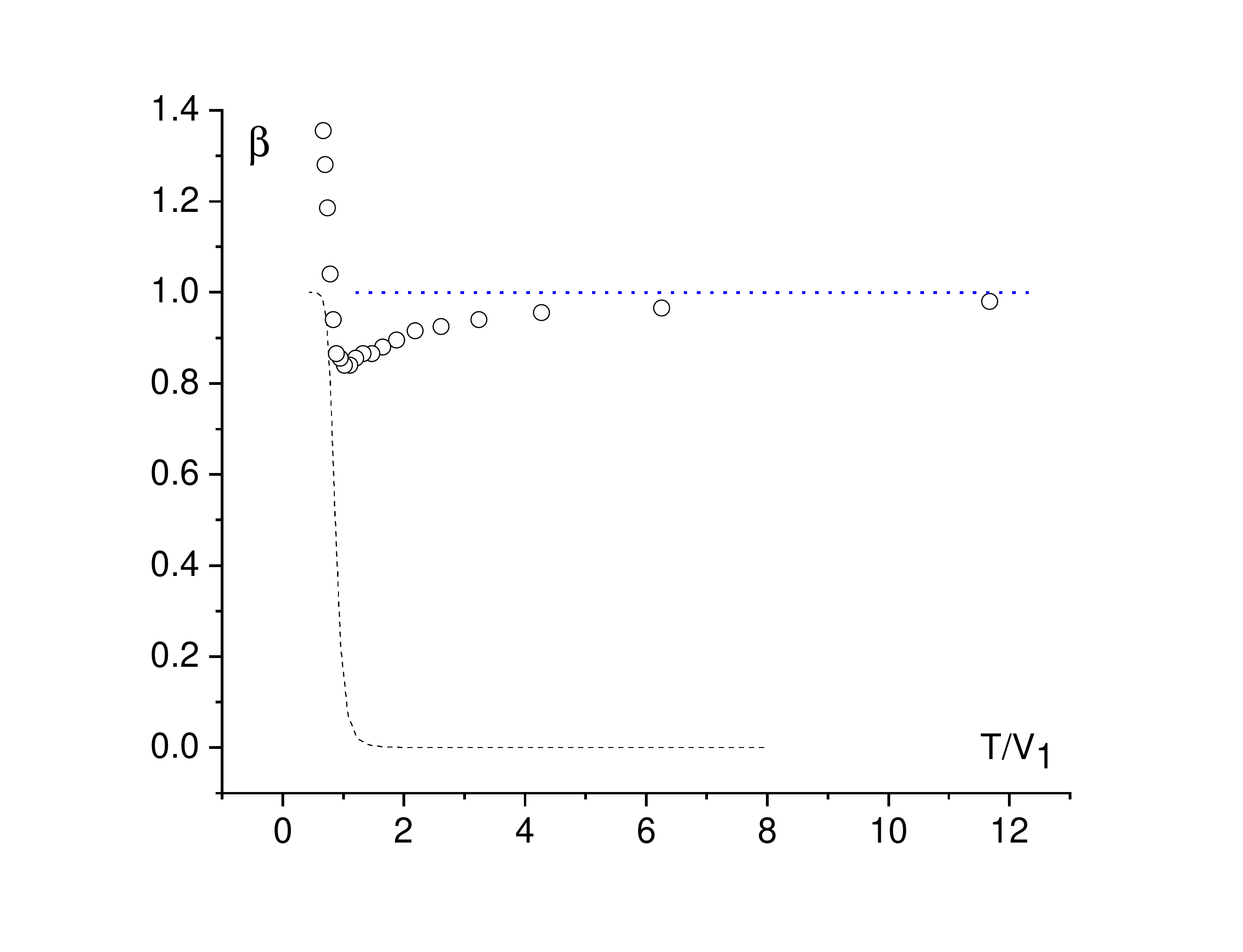}}}
\caption{Open circles show $\beta$ obtained from fitting the solution of Eq. \eqref{inter} ($V_0=0.3$, $V_1=0.115$) to \eqref{qfit}. The dashed line shows function $f$ in Eq. \eqref{inter}. The horizontal dotted line shows $\beta=1$.}
\label{fig2}
\end{figure}

We numerically integrate $q$ in Eq. \eqref{inter}, fit it to Eq. \eqref{qfit} and show the result in Fig. 3. We observe that, starting from high temperature, $\beta$ decreases from 1 to about 0.8 and up to the point where the fraction of SER processes remains dominant. At lower temperature, where CER processes dominate, $\beta$ starts increasing back to 1 and above. We therefore find that our model reproduces the experimental data of Ref. \cite{ruta}, where $\beta<1$ above the glass transition temperature $T_g$ but increases above 1 below $T_g$. The typical values of $\beta$ measured experimentally in \cite{ruta} are also reproduced. We note that the available experimental data \cite{ruta} do not indicate whether $\beta$ of CER reaches a plateau at low T. This remains an interesting open question for future work.

\section{Glassy relaxation: from slow dynamics to avalanches and Gardner physics}

Our model gives the following picture of relaxation in liquids and glasses. The SER phenomenon in the supercooled liquid state is due to the slow dynamics of local relaxation events (LREs). The activation energy barrier for LREs increases with time due to the elastic feed-forward mechanism, and is related to slow relaxation \cite{Micoulaut,Pineda,Egami_ageing} where the system moves down the energy landscape towards deeper minima. In the glass state below $T_g$, LREs proceed in an avalanche-like manner as seen experimentally~\cite{Evenson}, and the decrease of energy barriers for successively triggered LREs is related to the cascade-like propagation of LRE avalanches. This consistently explains why CER is experimentally observed in dynamical regimes of metallic glasses which are characterized by intermittent avalanche-like events~\cite{Evenson}. This picture also explains why CER is observed in metallic glasses and not in network glasses: the internal stresses in metallic glasses are much larger due to very fast cooling, leading to lower energy barriers for rearrangements as shown in~\cite{egami} and enabling the avalanche-like events. However, our model is general enough to predict CER in systems other than metallic glasses as long as these systems contain highly non-equilibrium states giving rise to avalanches.

Although the above picture is intuitive, the important point is that we derive both SER and CER on the basis of an atomistic model. This model also reproduces the CER-SER crossover observed experimentally across the glass transition~\cite{ruta}, and clarifies it as a two-state crossover between regimes with morphologically different energy landscape. The exact nature of this crossover and its connection to the Gardner transition, which also leads to a low-temperature regime of intermittent avalanche-like dynamics in the energy landscape~\cite{biroli}, are interesting questions for future work. The SER-CER crossover as described here appears as a promising candidate for the (so far elusive) experimental detection of the Gardner transition.

\section{Summary}

In summary, we presented an analytical atomistic model describing the recently observed compressed exponential relaxation in glasses. The model predicts the observed temperature variation of $\beta$ in both SER and CER regimes and the crossover between the two regimes, in agreement with recent experimental observations~\cite{ruta}. The model provides a broad picture of glassy dynamics and relaxation encompassing a slow relaxation above $T_g$ and an avalanche-dominated regime below $T_g$. A new connection between CER and the so-far elusive Gardner transition is proposed, to be further elucidated in future investigations.

\end{document}